\documentclass[journal,article,accept,moreauthors,pdftex,10pt,a4paper,universe]{mdpi} 
\usepackage{physics}
\usepackage{graphicx}
\usepackage{amsmath,amssymb,amsthm,dsfont,bm}
\usepackage{color}
\usepackage{soul}
\usepackage{cancel}
\usepackage{braket}
\firstpage{1} 
\articlenumber{x}
\doinum{10.3390/------}
\pubvolume{xx}
\pubyear{2016}
\copyrightyear{2016}
\externaleditor{Academic Editor: name}
\history{Received: date; Accepted: date; Published: date}

 \theoremstyle{mdpi}
 \newcounter{thm}
 \newcounter{ex}
 \newcounter{re}
\Title{Parameter estimation of wormholes beyond the Heisenberg limit}

\Author{Carlos Sanchidri\'an-Vaca $^{1,\dagger,\ddagger}$  and Carlos Sab\'in  $^{2,}$*}

\address{%
$^{1}$ \quad Departamento de F\'isica Te\'orica, Universidad Complutense de Madrid, Plaza de Ciencias, 1 28040 Madrid,Spain
; casanc05@ucm.es\\
$^{2}$ \quad Instituto de F\'isica Fundamental, CSIC,
Serrano, 113-bis,
28006 Madrid, Spain; csl@iff.csic.es}

\corres{Correspondence: casanc05@ucm.es}

\abstract{We propose to exploit the quantum properties of nonlinear media to estimate the parameters of massless wormholes. The spacetime curvature produces a change in length with respect to Minkowski spacetime that can be estimated in principle with an interferometer. We use quantum metrology techniques to show that the sensitivity is improved with nonlinear media and propose a nonlinear Mach-Zehnder interferometer to estimate the parameters of massless wormholes that scales beyond the Heisenberg limit. }


\begin{document}



\setcounter{section}{-1} 
\section{Introduction}
In last years, there have been many theoretical proposals to improve measurements through quantum technologies in the context of quantum metrology \cite{lloyd}, whose goal is to estimate  unknown parameters of interest. Among many other developments, such as gravitational wave detection \cite{squeezed} or timekeeping \cite{komar}, there have also been more challenging ideas, such as the measurement of acceleration with a Bose Einstein Condensate setup \cite{relativistic} or temperature through Berry's phase in \cite{martinez}.

Along these lines, it is natural to ask whether quantum metrology can be useful to test gravitational effects such as gravitational time delay (see, for instance, \cite{zych}). 
In particular, nonlinear media were proposed in \cite{kish} to estimate the Schwarzschild radius of the Earth beyond the Heisenberg limit. Moreover, the detection of distant traversable wormhole spacetimes by means of quantum metrology was considered in \cite{sabin}. Following this path, we propose to use the nonlinear Kerr effect to estimate the size of the throat of  distant massless wormholes with super-Heisenberg scaling. Given the renewed interest in the physics of wormholes coming from different fields \cite{cardoso,ringing,epr} and the proposal of new methods for their detection by classical means \cite{lensing}, our main goal is to explore the fundamental sensitivity limits provided by quantum mechanics, which in principle overcome the classical ones.

We consider the evolution of a coherent state under a nonlinear hamiltonian. If the propagation takes place in the spacetime of a distant traversable wormhole, the phase of the coherent state acquires a slight dependence on the relevant parameter of the metric, which in this case is the radius of the wormhole throat. The fundamental bound on the precision with respect to this parameter is given by the Quantum Fisher Information (QFI). Using the QFI, we find that the scaling with respect to the number of photons surpasses the Heisenberg limit, as expected, due to the nonlinearity. Moreover, we show that a realistic measurement protocol can get close to this fundamental limit by proposing a particular scheme of detection, 
consisting on an interferometer whose arms are stretched in a curved spacetime. Since we consider an almost-flat spacetime, we can express the metric as Minkowski plus a little perturbation, yielding a small correction to the length of the arm. This generates a phase shift in the interferometer that in principle can be measured.  We show that a standard homodyne detection scheme with a nonlinear interferometer would also exhibit super-Heisenberg scaling with respect to the number of photons.

In the linear regime, gravitational bodies produce a stationary phase shift which is locally indistinguishable from other sources. We find compulsory to rotate the interferometer in order to get a controlled oscillating phase shift.
The whole scheme resembles a laser interferometer setup for gravitational wave detection, such as LIGO. Indeed, we use experimental parameters from LIGO to show that parameter estimation might be, in principle, possible. Thus we also show that LIGO could in principle benefit from the use of nonlinear media. 

The structure of the paper is the following. In the first two sections we briefly recall some basic notions of quantum metrology and traversable wormholes, respectively. Then in the next section we present our results both for the QFI and the interferometric setup. Finally, we summarize our conclusions in the last section.

\section{Quantum Metrology}
Quantum metrology is the branch of physics which attempts to improve the estimation of parameters of interest through quantum resources. 
To infer the value of a parameter, $\theta$ from the data collected by n measurements $(x_1,x_2,...,x_n)$, we must build an estimator $\hat{\theta}$ that is a function of the possible outcomes.
The statistical error  is bounded by the classical Fisher information (FI), which represents the amount of information that a random variable x carries about an unknown parameter $\theta$ of a distribution that models x. It can be written as
\begin{equation}
F(\theta )=-\expval{\frac{\partial \log p(x;\theta )}{\partial \theta } }
\end{equation}
where $p(x;\theta )$ represents the probability that the outcome of a measurement is x when  the parameter is $\theta$.
Fisher information  can be extended to a metric on a statistical manifold and it obeys the Cramer-Rao inequality for estimating the variance of $\theta$.
\begin{equation}
\Delta  {\theta } \geq \frac {1}{\sqrt{F(\theta)}}
\end{equation}
In the quantum realm, $p(x;\theta )=\text{Tr}(\rho_\theta \Pi_x)$. 
The ultimate bound on the sensitivity $\Delta \theta_\rho$ of a state $\rho$ with respect to the parameter $\theta$  is obtained maximizing the FI over all possible measurements:
 \begin{equation}\label{QFIth}
\Delta \theta_\rho \geq {\frac {1}{\sqrt{\mathcal{H}(\theta)}}}
\end{equation}
where $\mathcal{H}(\theta)$ is the Quantum Fisher Information (QFI), which represents the maximum information that can be obtained.
For instance, in the case of pure states, we can estimate the QFI as $\mathcal{H}=4\Delta H$, where $H$ is the Hamiltonian. A possible loophole of this analysis is the assumption that the observed model is the predicted, so in case there is a different model underneath, we are estimating a false parameter.  However, Bayesian theory is probably the best way to deduce the  model that fits the data.

In quantum metrology, a typical parameter is the phase $\theta$ and a typical experimental set-up is an interferometer \cite{lloyd}.
It is a well-known result that the phase estimation for a coherent state with $N$ photons per pulse goes as $\Delta \theta \propto \frac{1}{\sqrt{N}}$, which is the Standard Quantum Limit (SQL) and is a consequence of the central limit theorem.
This resolution can be improved using quantum states such as squeezed light, achieving the so-called Heisenberg limit which set bounds to the sensitivity in the linear case as  $\Delta \theta \propto \frac{1}{N}$.
As it has been shown by different authors \cite{nonlinear1,nonlinear2,nonlinear3,nonlinear4}, the precision can be improved  with nonlinear media  as  $\Delta \theta \propto \frac{1}{\sqrt{N^3}}$. This limit has been confirmed experimentally \cite{nonlinearexp}. 

\section{Traversable wormholes}
Wormholes are a class of spacetime predicted by Einstein's equations with interesting properties.
They allow for spacetime travel and closed timelike curves, rising paradoxes about causality, which are not so when  quantum mechanics is considered \cite{closed}. 
The creation of a traversable wormhole would require some exotic source of negative energy, whose existence is  bounded by quantum energy inequalities (QIs) and  is dubious in classical physics.   The QIs entail that the more negative the energy density is in some time interval, the shorter its duration.
These constraints apply only to free quantum fields \cite{thoughts} and they extend to curved spacetime.
Intuitively,  light rays converge while coming into the wormhole throat but they will diverge once they leave so negative energy is compulsory for a timelike observer.
Moreover, it has been argued that a traversable wormhole must be only a little larger than Planck size, otherwise, its negative energy must be confined in a shell many orders of magnitude smaller than the throat size \cite{constraints}. However, there might be unexplored possibilities to avoid this bound, such as consider interacting quantum fields \cite{fewster}.
It remains an open question whether wormholes are forbidden by the laws of physics. 

Interestingly, wormholes can mimic black-hole metrics at first approximation, which in principle could question the identity of objects at the center of the galaxies as well as the observed gravitational waves \cite{cardoso},\cite{ringing}. Besides, a recent conjecture $ER=EPR$, involves wormholes and entanglement. It states that there might be difference kind of bridges for each kind of entanglement \cite{epr}. 

The estimation of parameters of these objects, as well as a general curved spacetime can be improved, in principle, by quantum metrology \cite{sabin}, which surpasses classical techniques such as gravitational lensing \cite{lensing}.

Let us consider a field propagating along the radial direction  in the  presence of a traversable  wormhole whose general metric is given by \cite{wormhole}
\begin{equation}
    ds^2=-c^2e^{2\Phi}dt^2+\frac{1}{1-\frac{b(r)}{r}}dr^2+r^2(d\theta^2+\sin^2\theta d\phi^2),
\end{equation} 
where $\Phi$, the redshift function and $b(r)$, the shape function are arbitrary functions of the radius r.
In the case of an Ellis massless wormhole, $\Phi=0$ and $b(r)=\frac{b_0^2}{r}$. Ignoring the angular part results in: 
\begin{equation}\label{metric}
ds^2=-c^2dt^2+\frac{1}{1-\frac{b_0^2}{r^2}}dr^2.
\end{equation}
As we will see below, for a far-away observer, the metric can be expressed as flat with a perturbation on the spatial part. Then, there will be a phase shift in the presence of a wormhole in the radial direction.

\section{Parameter estimation of wormholes through nonlinear effects}

The Kerr effect is produced when the polarization of the medium is proportional to $|E|^2$  and is well-known in nonlinear optics.  In quantum optics, it is usually used to generate non-classical states of light such as Schrodinger's cats.

Basically, there is a coupling between the intensity $I$ and the proper time $\tau$ in the phase \cite{coupling}:
\begin{equation}
\phi=\chi\tau I,
\end{equation}
where $\chi$ is the nonlinearity.
It has been shown that in the presence of a nonlinear medium the  phase can be detected with higher precision than the Heisenberg limit, replacing the vacuum by a nonlinear medium \cite{nonlinear1,nonlinear2,nonlinear3,nonlinear4,nonlinearexp}.
Classically, nonlinear effects can be interpreted as anharmonic terms in the Hamiltonian.
Then, we can consider the following quantum Hamiltonian:

\begin{equation}
\hat{H}= \chi (\hat{a}^\dagger \hat{a})^2 + \omega \hat{a}^\dagger \hat{a}= \chi \hat{n}(\hat{n}+1)+\hat{n}kc.
\end{equation}
where $a$ and $a^{\dagger}$ are the standard creation and annihilation operators of a single electromagnetic field mode. As we will see, the effective nonlinearity becomes $\chi\tau$, which essentially implies a coupling with the curvature of spacetime.

The more general form of a Gaussian state is a squeezed coherent state for a quantum harmonic oscillator, which is given by:
\begin{equation}
|\alpha ,\zeta \rangle =D(\alpha )S(\zeta )|0\rangle.
\end{equation}
A laser is the typical coherent source of the electromagnetic field, which is represented by a coherent state and is obtained by letting the unitary displacement operator D($\alpha$) act on the vacuum,
\begin{equation}
    |\alpha \rangle =e^{\alpha {\hat {a}}^{\dagger }-\alpha ^{*}{\hat {a}}}|0\rangle =D(\alpha )|0\rangle=e^{-{|\alpha |^{2} \over 2}}\sum _{n=0}^{\infty }{\alpha ^{n} \over {\sqrt {n!}}}|n\rangle .
\end{equation}
The evolution of the coherent state with frequency $\omega$ propagating through a nonlinear medium, is given by:

\begin{equation}
\ket{\alpha_{NL}(\tau)}=\hat{U}\ket{\alpha}= e^{-i\tau\hat{n}(\chi (\hat{n}+1)+kc)}\ket{\alpha} .
\end{equation}

It is impossible to give an analytic expression of the phase due to the nonlinear terms which produce a superposition of coherent states. However, we can give a bound for the variance of the phase.
\subsection{Quantum Fisher Information}

As explained before, the bound for the variance of an estimator is given by the Cramer-Rao inequality.  The Quantum Fisher Information for estimating the parameter $\tau$, $\mathcal{H}(\tau)$, is given by \cite{kish},

\begin{equation}
\mathcal{H}(\tau)=\lim_{d\tau\rightarrow 0}\frac{8[1-\sqrt{\mathcal{F}(\rho(\tau),\rho(\tau+d\tau)))]}}{d\tau^2} 
=4N\{[\omega+2(N+1)\chi]^2 +2N\chi^2\},
\end{equation}
where  $\mathcal{F}$ is the fidelity, $\mathcal{F}(\rho,\sigma)=[\text{Tr}\sqrt{\sqrt{\rho}\sigma\sqrt{\rho}}]^2$ which is a distance between two states $\rho, \sigma$ in the Hilbert space. When N, the number of photons per pulse, is large $N\rightarrow\infty$, the Heisenberg limit is surpassed 
\begin{equation}
    \Delta \tau \propto \frac{1}{N^{3/2}}.
\end{equation}
The proper time for a free-falling observer has a dependence on $b_0$, so the quantum Fisher information may be rewritten as
\begin{equation}\label{b0}
    \mathcal{H}(b_0)=\left|\frac{\partial \tau}{\partial b_0}\right|^2\mathcal{H}(\tau)
\end{equation}

In 2D , the metric \eqref{metric} becomes flat  by means of a change of coordinates:
\begin{equation}
ds^2=-c^2dt^2+dl^2,
\end{equation}
where
\begin{equation}
l=\sqrt{r^2-b_0^2}.
\end{equation}
Nevertheless, there is a phase shift for propagation in the radial direction. To see this, notice that for a free-falling observer, the proper length equals the proper time times $c$, so it can be expressed in laboratory coordinates as: \begin{equation}\label{proper}
   c\tau= L'=|l_2-l_1| =\left|\sqrt{r_2^2-b_0^2}-\sqrt{r_1^2-b_0^2}\right|.
\end{equation}
We assume that the separation $L'=c\tau$ is small as compared to the  distance to the wormhole;  $L<<$ $r_1,r_2$, and for simplicity we consider that $r_2 > r_1$, thus $L= r_2 - r_1$.
If the propagation takes place far away from the wormhole throat, $r_1, r_2 >>b_0$,  the proper length can be approximated using a Taylor expansion up to second order in $\frac{b_0^2}{r_1^2}$, giving
\begin{equation}
    \tau\approx L/c\left(1+\frac{b_0^2}{2r_1^2}-\frac{b_0^2}{2r_1^2}\frac{L}{r_1}\right).
\end{equation}
So the proper length as measured in the radial direction of the Ellis metric is larger than the Minkowski one.
Now, we calculate the Jacobian for estimating the QFI as a function of $b_0$ in \eqref{b0}:
\begin{equation}\label{jacobian}
\frac{\partial \tau}{\partial b_0}=L \left(\frac{b_0}{cr_1^2}-\frac{b_0L}{cr_1^3}\right).
\end{equation}
Finally, putting everything together the sensitivity is given by:
\begin{equation}\label{nonlinear}
\begin{split}
&\frac{\Delta b_0}{b_0}
=\frac{1}{b_0}\frac{1}{\sqrt{\mathcal{H}(b_0)}}=\frac{1}{b_0}\left|\frac{\partial b_0}{\partial \tau} \right|\frac{1}{\sqrt{\mathcal{H}(\tau)}}=\\
&\simeq \frac{1}{L(\frac{b_0^2}{cr_1^2}-\frac{Lb_0^2}{cr_1^3})}\frac{1}{\sqrt{N\{[\omega+2(N+1)\chi]^2 +2N\chi^2\}}}\\
&\simeq\frac{1 }{ L (\frac{b_0^2}{2r_1^2}-{\frac{Lb_0^2}{2r_1^3}})}\frac{1}{\sqrt{N\{[\omega+2(N+1)\chi]^2 +2N\chi^2\}}}\\
&\simeq\frac{cr_1^2}{2Lb_0^2}\frac{1}{\sqrt{N\{[\omega+2(N+1)\chi]^2 +2N\chi^2\}}}.
\end{split}
\end{equation}

In the last step we have neglected the term $\frac{b_0^2}{2r_1^2}\frac{L}{r_1}$ and expanded in Taylor series around 0 in x and y the ratio $1/(x+xy) \approx \frac{1}{x}$, where  $x=\frac{b_0^2}{2r_1^2}$ and $y=-\frac{L}{r_1}$.
Therefore, when N goes to infinity we find a polynomial enhancement in the sensitivity which goes beyond the Heisenberg limit;
\begin{equation}
\frac{\Delta b_0}{b_0} \propto \frac{1}{N^{3/2}}.
\end{equation}
Moreover, repeating the experiment M times, we would reduce the noise as a Gaussian variance:
\begin{equation}\label{nonlinearbis}
   \frac{\Delta b_0}{b_0}=\frac{c r_1^2}{2Lb_0^2}\frac{1}{\sqrt{NM\{[\omega+2(N+1)\chi]^2 +2N\chi^2\}}}.
\end{equation}
Note that in the limit $\chi=0$ we recover the results in \cite{sabin}.
As usual in quantum metrology, with the QFI we have the best theoretical value of the sensitivity. However, we do not know yet whether there is a realistic measurement protocol achieving this limit. This is the goal of the next subsection.

\subsection{Nonlinear interferometer}

\begin{figure}[h]
    \includegraphics[scale=0.7]{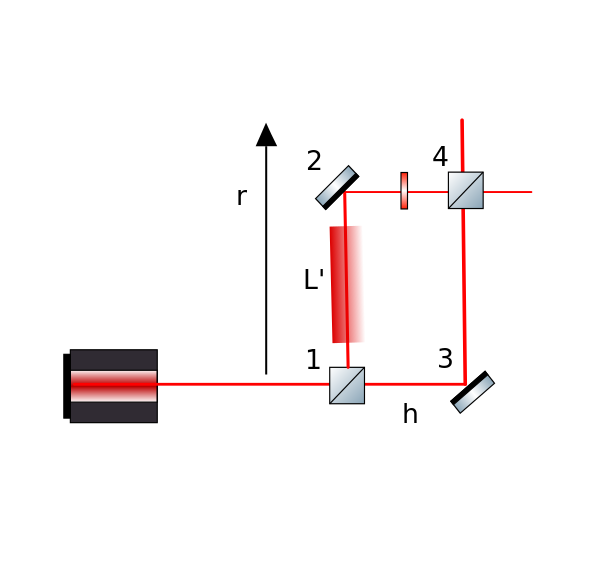}
    \caption{A Mach-Zehnder interferometer with a non linear medium placed along the radial direction.}
    \label{interferometer}
\end{figure}

Let us consider a more realistic scenario in which the measurement is carried out by homodyne detection. This consists in estimating an unknown signal comparing it with a known local oscillator.
The unknown signal is one of the output modes $b_k$, where k makes reference to the wave number in case there are several frequencies. As we assume a monochromatic source this will be suppressed later.
We follow the nonlinear Mach-Zehnder interferometer  proposed in \cite{kish} (see Figure \ref{interferometer}). Metrics are qualitatively different so the set up is not obtained just by replacing the black hole with a wormhole. We have:
\begin{equation}
b_k=\frac{1}{2}[a_k(e^{-ik(\phi_{12}+\phi_{24})}-e^{-ik(\phi_{13}+\phi_{34})})] \\
 +v_k(e^{-ik(\phi_{12}+\phi_{24})}+e^{-ik(\phi_{13}+\phi_{34})}),
 \end{equation}
where $a_k$ is an input coherent state and $v_k$ an input vacuum state. 
We must introduce only a non linear material in one intereferometer arm and an adjustable linear phase $\beta$ in the other arm.

As we will explain in the next section, we should rotate the interferometer in order  to determine the position of the wormhole. 
The proper distance is only modified basically in the radial axis, this will produce an oscillating signal that allows to identify the radial axis of the wormhole where the phase shift is maximum.
In fact, the proper distance on the top and bottom arms will differ but we can neglect this term with respect to the main correction in the radial direction as will be explained in detail below.

We assume that the length of the top and bottom arms, $h$, is small with respect to L $(h<<L)$, so we can neglect curvature effects and take the value $\phi_{13}=\phi_{24}=0$. Using that $c\tau=L'$, it follows that $\phi_{34}=0$ so the vertical output mode  turns out to be
\begin{equation}
b_k=\frac{1}{2}(e^{-ik\phi+i\chi \tau a_k^\dagger a_k}
    -e^{i\beta})a_k+\frac{1}{2}(e^{-ik\phi+i\chi \tau a_k^\dagger a_k}+e^{i\beta})v_k.
\end{equation}
The phase shift is induced in one arm of the interferometer through a nonlinear medium such that
\begin{equation}
\phi=\phi_{12}=L'-\frac{c}{n'}\tau\simeq L\left(1+\frac{b_0^2}{2r^2}\right)\left(1-\frac{1}{n'}\right), 
\end{equation}
where 
\begin{equation}
    \tau\approx \frac{L}{c} \left(1+\frac{b_0^2}{2r^2}\right)
\end{equation}  is the proper time, $L'$ the proper distance, $n'$ is the first-order refractive index of the material, (we consider the generic value  $n'=1.5$) and $r$ is the distance to the wormhole previously called $r_1$.

We use the approximation of linearized Gaussian regime given in \cite{kish} valid when $\tau\chi\sqrt{N}<<1$ so that

\begin{equation}
    e^{-i\chi\tau a_k^\dagger a_k}\approx e^{-i\chi|\alpha|^2\tau}[1-i\chi\tau(\alpha^*\delta a +\alpha \delta a^\dagger)]\alpha+e^{-i\chi|\alpha|^2\tau}\delta \alpha
\end{equation} 
\begin{figure}[h]
    \centering
    \includegraphics[scale=0.75]{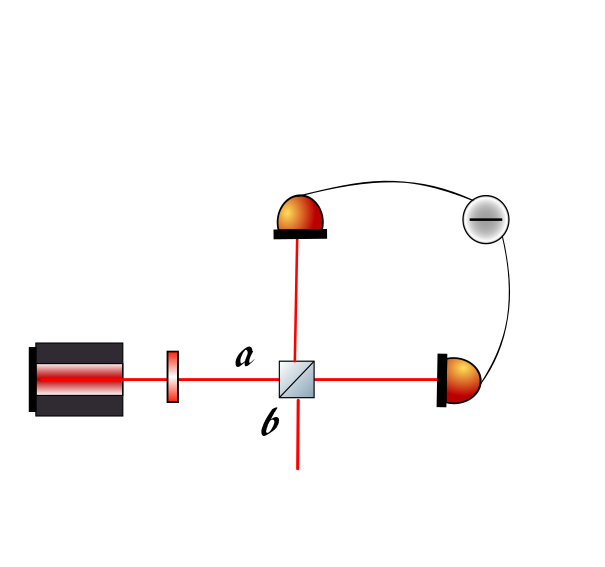}
    \caption{Homodyne detection: The unknown signal, $b$, is introduced into a beam splitter with a local oscillator, $a$,  whose phase $\theta$ is controlled. The difference of photons is proportional to the quadrature $X_b$.}
    \label{homodyne}
\end{figure}
The quadrature  $X_b=b(\tau)e^{i\theta} +b^\dagger(\tau)e^{-i\theta}$ can be measured by homodyne detection as can be seen in Figure \ref{homodyne}. This can be written as
\begin{eqnarray}
X_b&=&|\alpha|[\cos(\theta+\xi)-\cos(\theta+\beta)]
    -\chi|\alpha|^2\tau[\sin(\theta+\xi)-\sin(\theta+\beta)]X
   \nonumber\\ &+&\frac{1}{2}(X_{\theta+\xi}-X_{\theta+\beta})+\frac{1}{2}(X^v_{\theta+\xi}-X^v_{\theta+\beta} )
\end{eqnarray}
 where we have defined 
 \begin{equation}
\xi=k\phi-\tau\chi|\alpha|^2\simeq L\left(1+\frac{b_0^2}{2r^2}\right)\left( k\left(1-\frac{1}{n'}\right) - \frac{\chi|\alpha|^2}{c}\right)    
\end{equation}
Then, the expectation value turns out to be
\begin{equation}
    \expval{X_b}=|\alpha|(\cos(\theta+\xi)-\cos (\theta+\beta))
\end{equation}
and
\begin{eqnarray}
     \expval{\Delta X_b^2}&=&\chi^2|\alpha|^4\tau(\sin(\theta+\xi)-\sin(\theta+\beta))^2\nonumber\\
     &-&\chi|\alpha|^2\tau(\sin(\theta+\xi)-\sin(\theta+\beta))\cdot
     [\cos(\theta+\xi)-\cos(\theta+\beta)]+1
\end{eqnarray}
where we have used that for coherent states, $\expval{X}=0$, $\expval{X^2}=\frac{1}{2}$ and consequently $\expval{\Delta X^2}=\frac{1}{2}$. 
Now, defining $\gamma=\frac{b_0^2}{2r^2}$ we find that the derivative of X with respect the parameter is
\begin{equation}\label{derivative}
    \frac{d \expval{X_b}}{db_0}=2|\alpha|\left({kc}\left(1-\frac{1}{n'}\right) +|\alpha|^2\chi\right)\frac{2L}{cb_0}\gamma\sin(\theta+\xi).
\end{equation}
The nonlinearity creates noise from antisqueezing in the axis of rotation, which can be removed  making $\beta=\xi$, making the variance shot noise ($\expval{\Delta X_b^2}=1$).

In order to calculate the sensitivity of $b_0$ we use the sensitivity of the quadrature through the following equation
\begin{equation}
    \frac{\Delta b_0^2}{b_0^2}=\frac{\Delta X_b^2}{b_0^2\left| \frac{d \expval{X_b}}{db_0}\right|^2}.
\end{equation}
We can optimize our choice of $\theta$ in \eqref{derivative} making  $\theta =\pi/2-\xi$ which gives 
\begin{equation}
    \frac{d \expval{X_b}}{db_0}=\frac{4\gamma L}{cb_0}|\alpha|\left({kc}\left(1-\frac{1}{n'}\right) +|\alpha|^2\chi\right). 
\end{equation}
Finally, the sensitivity is:
\begin{equation}
\frac{\Delta b_0}{b_0}=\frac{c}{4\gamma L}\frac{1}{\sqrt{N[\omega\left(1-\frac{1}{n'}\right)+N\chi]^2 }}\\
=\left(\frac{cr^2}{L 2b_0^2}\right)\frac{1}{\sqrt{N[\omega\left(1-\frac{1}{n'}\right)+N\chi]^2 }}.\label{eq:sensbound}
\end{equation}
This expression is quite different from \eqref{nonlinear}, however, it still scales as $\frac{1}{N^{3/2}}$ (see Figure \ref{SQL}). We compare both expressions in figure \ref{quadrature}. 

\subsection{Parameter estimation}
\begin{figure}[h]
    \includegraphics[scale=0.90]{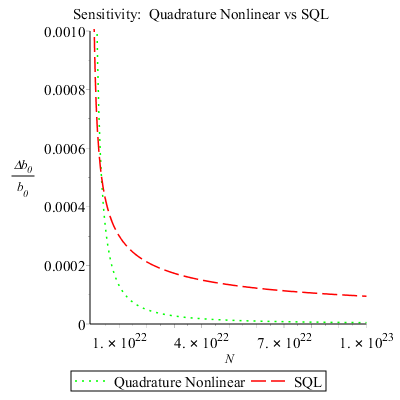}
    \caption{Sensitivity of the Quadrature Measurement with a nonlinear medium (Eq.(\ref{eq:sensbound})) together with theoretical SQL vs. number of photons $N$ for $L=1 \operatorname{km}$, $\omega= 10^{15} \operatorname{Hz}$, $\frac{r}{b_0}=10^8$, $n'=1.5$ and $\chi=10^{-8} \operatorname{Hz}$. We see that in the high-number regime shown in the plot, the sensitivity goes in principle significantly beyond the standard quantum limit. This range for the number of photons would correspond to a laser with a wavelength around a thousand of $\operatorname{nm}$ and a few $\operatorname{W}$ of power, as in LIGO.}
    \label{SQL}
 \end{figure}

\begin{figure}[h]
    \includegraphics[scale=0.90]{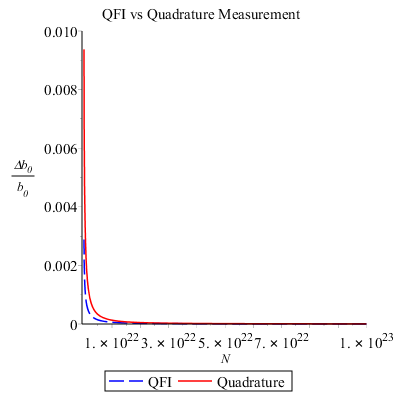}
    \caption{Theoretical sensitivity for the Quantum Fisher Information and Quadrature Measurement in the nonlinear cases, with the same parameters as in Figure \ref{SQL}. We see that in the high-number regime discussed in Figure \ref{SQL} our realistic measurement protocol achieves a precision extremely close to the fundamental limit, exhibiting super-Heisenberg scaling.}
    \label{quadrature}
\end{figure}

We now discuss the theoretical prospects for wormhole parameter estimation with these techniques. The nonlinear parameter $\chi$ depends on the third -order susceptibility $\chi^{(3)}$ because the second-order susceptibility $\chi^{(2)}$ vanishes due to inversion symmetry of the lattice.
Let us consider typical values of the LIGO interferometer (L=1 km), with frequency $\omega\sim 10^{15}$ Hz,  and a typical value of the nonlinear parameter $\chi=\frac{\hbar \omega^2 \chi^{(3)}}{\epsilon_0^2 V}\approx\frac{10^{-34}10^{30}10^{-20}}{10^{-24}}\approx 1$ Hz \cite{coupling}. We will consider several values in the range $\chi=1-10^{-8}$ Hz.
For the  ratio of $\frac{r}{b_0}=10^8$ and $\chi=10^{-8}$  Hz, sensitivity can be improved with a nonlinear interferometer by 2 orders of magnitude, as shown in Figure \ref{SQL},with respect to the theoretical SQL, which is obtained by making  $\chi=0$ in the QFI.
The sensitivity values for the theoretical QFI in \eqref{nonlinear} and the quadrature measurement in the nonlinear case are practically indistinguishable beyond $10^{22}$ photons as can be seen in figure \ref{quadrature}.

To analyze these results, we must set the range of parameters in which a wormhole can be expected in principle. Considering that the maximum tolerance is $\frac{\Delta b_0}{b_0}=0.1$, we can admit ratios of $\frac{r}{b_0} =10^{11}$. The size of a wormhole in the centre of the Milky way ( $10^4$ Pc), taking into account that $1  \operatorname{Pc}=3\cdot 10^{16} \operatorname {m}$,   would be $10^9 \operatorname{m}$, which seems too large. 
In the case of a black-hole mimicker, the throat is of the order of the Schwarzschild radius of the black hole, which in the case of LIGO is $200 \operatorname {km}$, detection would be possible from $r\simeq 1 \operatorname{Pc}$. 
Taking into account that $\frac{\Delta b_0}{b_0}\propto \frac{1}{L\chi}$, we can keep the same sensitivity for the same ratio $\frac{r}{b_0}$ if the distance is reduced while the non linear coupling is raised.
In fact, $\chi$ is expected to be bigger than $10^{-8} \operatorname{Hz}$ so the sensitivity could be improved significantly as shown in  \cite{experiment}.

We can consider a  modest value for the length of  the fiber  approximately given by $1 \operatorname{m}$. In this case it would be needed a larger nonlinear parameter ($\chi\simeq 10^{-5}  \operatorname{Hz}$) in order to obtain a ratio $\frac{r}{b_0}=10^8$. Including  repetitive measurements ($M=1 \operatorname{GHz} = 10^9 \operatorname{Hz}$), the ratio $\frac{r}{b_0}$ can be as high as $10^{12}$.
Considering $\chi\simeq 1 \operatorname{Hz}$, which has been reported experimentally, \cite{experiment}, the sensitivity improves by several orders of magnitude. A possible wormhole as small as $1 \operatorname{m}$ could be estimated at a distance of  $10 \operatorname {Pc}$.

\begin{figure}[h]
\includegraphics[scale=0.90]{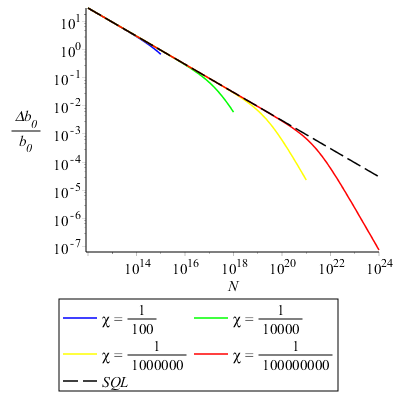}
\caption{Improvement of sensitivity with respect to the SQL vs. number of photons $N$ for several values of the nonlinear parameter, $L=1 \operatorname{m}$, $M=1 \operatorname{GHz} = 10^9 \operatorname{Hz}$ and $\frac{r}{b_0}=10^9$, with the rest of the parameters as in Figure \ref{SQL}. The nonlinear case becomes relevant for $\chi<10^{-4} \operatorname{Hz}$ in the high-number regime.}
    \label{improve}
\end{figure}

We plot results for several values of the nonlinear parameter and a ratio  $\frac{r}{b_0}=10^9$ with repetitive measurements and a value of $L=1 \operatorname{m}$ in Figure \ref{improve}.

We can also show that in the presence of optical losses or thermal noise, sensitivity will not vary substantially. Optical losses include a factor $\eta$ in the QFI \cite{losses} while for a thermal state  with an average number of excitations $n_T$, the QFI  will be reduced by a factor $ (1+2n_T)^{-1}$ \cite{thermal}:
\begin{equation}
\frac{\Delta b_0}{b_0}=\frac{\eta}{ 1+2n_T}\frac{cr_1^2}{2 Lb_0^2}\frac{1}{\sqrt{N[\omega\left(1-\frac{1}{n'}\right)+N\chi]^2 }}
\end{equation}
Further analysis of errors would depend on the particular experimental set up.

\subsection{Rotation and translation of the interferometer}

We have considered to be in an almost-flat space time with a tiny perturbation which does not depend on time. This might rise the question of how to actually determine the Minkoswki length, $L$.
For this, we need to rotate the interferometer two angles $\theta$ and $\phi$ around two independent axis. This would change the proper length producing an oscillating phase-shift, necessary for deducing the position of the wormhole. 

The interferometer must be translated along the radial axis so we can repeat the experiment and check the distance to the wormhole.

Assuming that the interferometer is in the plane $\phi=0$, and in the radial direction as desired, we can rotate the interferometer an angle $\theta$, resulting in the following change of coordinates:
\begin{equation}
\vec{e_r}=\vec{e_x}\rightarrow \cos \theta \vec{e_x}+\sin\theta \vec{e_y}.
\end{equation}
Notice that the module of the proper length would not change in the Minkowski metric due to Lorentz covariance, but this only apply locally. When rotating, the laboratory stops being an inertial reference frame. The proper length changes at first order as:
\begin{equation}
L_{\theta}=\sqrt{(L'\cos\theta)^2+(L\sin\theta)^2}\simeq\sqrt{L^2+\frac{L^2b^2_0}{r^2}\cos^2 \theta}=L(1+\frac{b^2_0}{r^2}\cos^2\theta ).
\end{equation}

\section{Conclusions}

We have applied quantum metrology techniques to estimate gravitational parameters. In particular, we propose to use nonlinear media for the detection of Ellis wormholes. The tiny perturbation generated by a distant wormhole on a flat spacetime induces a slight phase shift in a coherent state which propagates along the radial direction. We show that, in principle, this effect can be estimated with a precision which not only goes beyond the classical limits but can also overcome the Heisenberg limit. We show that the fundamental sensitivity limit provided by the QFI is not restricted by the Heisenberg limit when a nonlinear medium is considered. While this is an abstract result, we also show that a particular measurement protocol can get extremely close to the theoretical bound. In particular, we find that an standard homodyne detection protocol with a nonlinearity in one interferometer arm also exhibits super-Heisenberg scaling. We consider parameters of laser interferometers such as LIGO and realistic values of the nonlinearity, and discuss the theoretical prospects for parameter estimation. Of course, a complete experimental proposal would require an extremely careful characterization of all the sources of error -as in LIGO- and in any case would presumably be highly challenging. However, our goal is only to show that, in principle, an extremely sensitive estimation could be achieved by exploiting quantum resources. Along this vein, our results can motivate further exploration of this research path.

\vspace{6pt} 


\acknowledgments{C. S. has received financial support through the Junior Leader Postdoctoral Fellowship Programme from “la Caixa” Banking Foundation and Fundaci\'on General CSIC (ComFuturo Programme)}

\authorcontributions{C. S. proposed the general idea and supervised the work and the writing of the mansucript. C.S-V. developed the ideas, made all the computations and wrote the mansucript.}

\conflictofinterests{The authors declare no conflict of interest.} 




\bibliographystyle{mdpi}

\renewcommand\bibname{References}



\end{document}